\documentclass[epj]{svjour}
\usepackage{graphics}

\begin{document}

\title{Geometry effects in the magnetoconductance of normal and Andreev Sinai billiards}

\author{Nikolaos G. Fytas\inst{1}
}                     

\institute{Applied Mathematics Research Centre, Coventry
University, Coventry CV1 5FB, United Kingdom}

\date{Received: date / Revised version: date}

\abstract{We study the transport properties of low-energy
(quasi)particles ballistically traversing normal and Andreev
two-dimensional open cavities with a Sinai-billiard shape. We
consider four different geometrical setups and focus on the
dependence of transport on the strength of an applied magnetic
field. By solving the classical equations of motion for each setup
we calculate the magnetoconductance in terms of transmission and
reflection coefficients for both the normal and Andreev versions
of the billiard, calculating in the latter the critical field
value above which the outgoing current of holes becomes zero.
}

\authorrunning{Nikolaos G. Fytas} \titlerunning{Geometry effects in the magnetoconductance of normal and Andreev Sinai billiards}

\maketitle

\begin{figure}
\resizebox{1 \columnwidth}{!}{\includegraphics{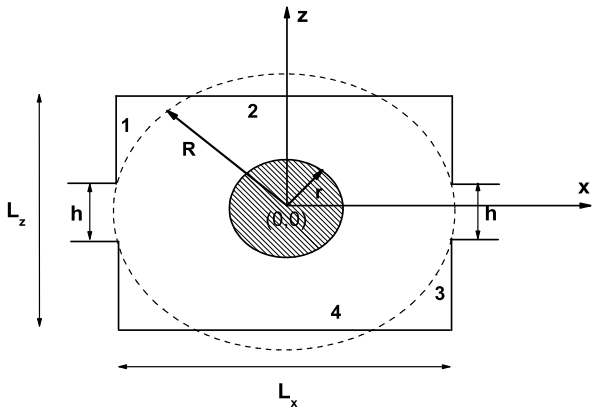}}
\caption{The general open geometry of the Sinai billiard
considered in this work.} \label{fig:1}
\end{figure}

Ballistic transport of particles across billiards is a field of
major importance due to its fundamental properties as well as
physical applications~\cite{QuaCh,Alha,Ri00,Jalab}. In such
systems, a two-dimensional cavity is defined by a step-like
single-particle potential where confined particles can propagate
freely between bounces at the billiard walls. For open systems the
possibility of particles being injected and escaping through holes
in the boundary is also allowed. As an example, we consider the
open geometry of the extensively studied Sinai billiard shown in
figure~\ref{fig:1}. Experimental realizations are based on
exploiting the analogy between quantum and wave mechanics in
either microwave and acoustic cavities or vibrating
plates~\cite{QuaCh}, and on structured two-dimensional electron
gases in artificially tailored semiconductor
heterostructures~\cite{Alha,Ri00,Jalab}. In the latter case, the
particles are also charge carriers making these nanostructures
relevant to applied electronics.

\begin{figure}
\resizebox{0.6 \columnwidth}{!}{\includegraphics{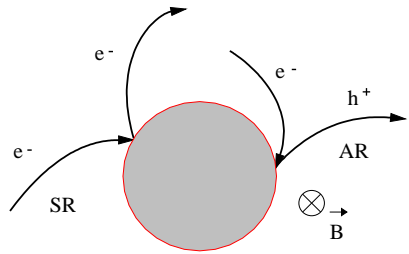}}
\caption{(color online) Typical specular (SR) and Andreev
reflection (AR) at the circular central antidot of
figure~\ref{fig:1}. A magnetic field is applied as indicated.}
\label{fig:2}
\end{figure}

Focusing the attention on the electronic analogues, more recently
the possibility to couple a superconductor to a ballistic quantum
dot has been considered both
theoretically~\cite{KosMasGol,S-billiards} and
experimentally~\cite{EPL02ETW}, so that some part of the billiard
boundary exerts the additional property of Andreev
reflection~\cite{And64}. During this process particles with
energies much smaller than the superconducting gap are coherently
scattered from the superconducting interface as Fermi sea holes
back to the normal conducting system (and vice versa).
Classically, Andreev reflection manifests itself by
retroreflection, i.e., all velocity components are inverted,
compared to the specular reflection where only the boundary normal
component of the velocity is inverted. Thus, Andreev reflected
particles (holes) retrace their trajectories as holes (particles).
If, however, a perpendicular magnetic field is applied in
addition, such retracing no longer occurs due to the inversion of
both the charge and the effective mass of the quasiparticle
resulting in opposite bending. Typical trajectories are
illustrated in figure~\ref{fig:2}.

A unique feature of this class of (quantum) mechanical systems is
their suitability for studying the quantum-to-classical
correspondence. In particular, much effort has been devoted in
revealing the quantum fingerprints of the classical dynamics which
may be parametrically tuned from regular to chaotic via, e.g.,
changes in the billiard-shape. A range of theoretical tools has
been used, spanning the usual analysis of classical trajectories
and the semiclassical approximation to the models of Random Matrix
Theory and fully quantum mechanical calculations. The main
signatures of classical integrability (or lack of it) on the
statistics of energy levels and properties of the transport
coefficients for closed and open systems, respectively, have been
discussed in detail in various
reviews~\cite{QuaCh,Alha,Ri00,Jalab}. Discussions on modifications
owing to the possibility of Andreev reflection appear in more
recent
studies~\cite{KosMasGol,S-billiards,MBFC96,PRB00CBA,EPJB01ILV,PRL99SB,TA01,PRB04CPP,FTPR05},
mostly focusing on the features of the quantum mechanical level
density.

The validity of classical calculations of these type of systems
has been revealed in reference~\cite{F05}. In fact, it has been
shown that a purely classical analysis may provide qualitative
rationalization and quantitative predictions for the average
quantum mechanical transport properties of a generic billiard,
such as the square cavity shape of figure~\ref{fig:1}, both in the
presence or absence of Andreev reflection. Moreover, while in most
previous works only the cases of zero or small magnetic field have
been considered, in reference~\cite{F05} the regime of finite
magnetic field strengths has been analyzed and it has been shown
that the classical trajectories, that depend parametrically on the
applied magnetic field, suffice to describe the overall features
of the observed non-monotonic behavior. Within this viewpoint, in
the present work we study classically the ballistic transport of
charge carriers across different geometrical setups originating
from the general form of the Sinai-billiard shown in
figure~\ref{fig:1}, under an externally applied magnetic field
$B$. We consider four different setups ($W$ is our scaling unit in
what follows): (a) a square cavity - centered antidot (sc) setup
for which $L_{\rm x} = L_{\rm z} = 5W$ and $h = r = W$ ; (b) a
square cavity - displaced antidot (sd) setup where the geometric
scaling follows setup (a) but now the center of the antidot is
displaced at ($r,0$) ; (c) a rectangular cavity - centered antidot
(rc) setup where $L_{\rm x} = 5W$, $L_{\rm z} = 3.75W$, and $h = r
= W$ ; and finally (d) a circular cavity - centered antidot (cc)
setup with $R = 2.5W$ and $h = r = W$ as before. In this
particular setup (d) the cavity is shown by the dashed circle in
figure~\ref{fig:1}. In all cases the symmetric leads attached to
the left and right side of the cavity define source and sinks of
quasiparticles. Note that, the central scattering disk can be
either a normal or a superconducting antidot. In the former case
the antidot represents an infinitely high potential barrier while
in the latter case it is considered as an extended homogeneous
superconductor characterized by the property of Andreev
reflection~\cite{KosMasGol}. Experimentally, such antidot
structures have been realized in periodic arrangements, thus
forming superlattices~\cite{Ri00,EPL02ETW}. The boundaries of the
square and rectangular cavity, numbered clockwise by the labels 1
through 4 in figure~\ref{fig:1}, are always normal conducting
potential walls of infinite height. The same applies also for the
case of the circular cavity, and in particular for the upper and
lower semicircles (dashed lines in figure~\ref{fig:1}).

The general form of the Hamiltonian describing the dynamics of
charged particles inside the cavity reads
\begin{equation}
\mathcal{H}=\frac{1}{2 m^\ast_\alpha}(\vec{p}-q_\alpha \vec{A})^2.
\label{eq:hamiltonian}
\end{equation}
The index $\alpha$ is used to describe the possibility that the
propagating particles are either electrons ($\rm e$) or holes
($\rm h$). This generalization is necessary for a correct
description of the dynamics in the setup with the superconducting
antidot. The canonical momentum vector is $\vec{p}=(p_x,p_z)=
m^*_\alpha \vec{v} + q_\alpha \vec{A}$ where $\vec{v}$ is the
mechanical velocity, the corresponding position vector being
$\vec{r}=(x,z)$. Charge conservation yields $m^*_h=-m^*_e$ for the
effective masses and $q_{\rm h}=-q_{\rm e}$ for the electric
charge. The main property which distinguishes the two cases, i.e.,
normal/superconducting antidot, is the interaction of the charged
particle with the scattering disk. The latter is captured by the
elementary processes illustrated in figure~\ref{fig:2}, namely,
specular reflection (SR) versus the Andreev reflection (AR).

\begin{figure}
\begin{center}
\resizebox{1 \columnwidth}{!}{\includegraphics{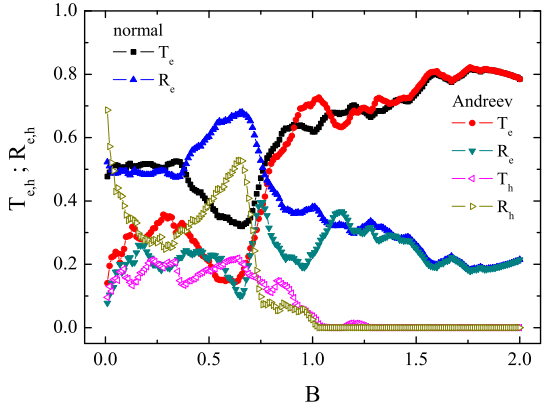}}
\caption{(color online) Magnetic-field dependence of the
transmission and reflection coefficients for the normal and
Andreev version of the setup (b) of the Sinai billiard of
figure~\ref{fig:1}. The field strength is in units of $B_0$.}
\label{fig:3}
\end{center}
\end{figure}

The initial conditions for incoming electrons are determined by
the phase-space density $\rho_o(x,z,v_x,v_z) = \frac{1}{2 m^\ast_e
v W}\delta(x+\frac{L}{2})\times
  \left[\Theta(z+\frac{W}{2})-\Theta(z-\frac{W}{2})\right]\times
\delta(m^*_e(v-v_F))\cos \theta$, where $\theta \in
[-\frac{\pi}{2},\frac{\pi}{2}]$ is the angle of the initial
electron momentum with the $x$-axis and $v_{\rm F}=\sqrt{2 E_{\rm
F}/m^{\ast}_{\rm e}}$ and the coordinate origin is assumed at the
center of the cavity. The trajectories of the charged particles in
the billiard consist of segments of circles with cyclotron radius
$r=m^\ast_\alpha v/ (-q_e B)$ (with $v=\sqrt{v_x^2+v_z^2}$). For
the magnetic field the symmetric gauge
$\vec{A}=[(B/2)z,0,-(B/2)x]$ has been chosen, accounting for a
homogeneous magnetic field of strength $B$ in $y$-direction,
perpendicular to the two-dimensional system. In what follows, we
define as magnetic field unit the value $B_{0}=(m_{\rm
e}^{\ast}v_{\rm F})/(-q_{\rm e}W)$ for which the cyclotron radius
is equal to $W$. It is convenient to use a dimensionless form of
the classical equations of motion by employing the scaling
$x=\xi_x W$ and $z=\xi_z W$ for the spatial coordinates and
$t=\tau / \omega $ (with $\omega = B_0/m^*_e$) for the time
coordinate. The above quantities are calculated for 100 values of
the magnetic field strength varying from 0.01 to 2 using an
ensemble of $10^6$ different initial conditions distributed
according to the phase-space density $\rho_o(x,z,v_x,v_z)$ given
above for each $B$-field value. The magnetic field dependence of
typical transmission and reflection coefficients for electrons and
holes $T_{\rm e, \rm h}$ and $R_{\rm e, \rm h}$ respectively is
shown in figure~\ref{fig:3} for the case of the square cavity with
a displaced antidot [setup (b) of the Sinai billiard of
figure~\ref{fig:1}] for both a normal and a superconducting
(Andreev) version. The obtained curves (similarly also for the
other setups) are quite irregular, possibly indicating the
presence of fractal fluctuations in the magnetoconductance of the
system~\cite{Ketz}, as will also be seen below.

\begin{center}
\begin{figure}
\resizebox{1 \columnwidth}{!}{\includegraphics{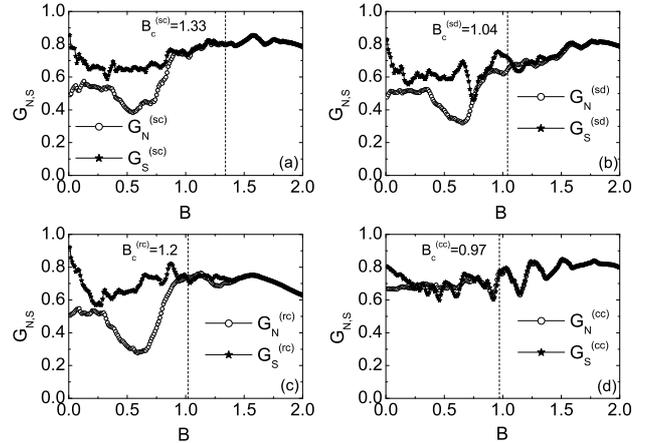}}
\caption{Magnetoconductance for the four different setups
discussed in the text. In each case results for both the normal
(open circles) and Andreev (filled stars) version of the Sinai
billiard is shown. The dotted line corresponds to the critical
field value above which the outgoing current of holes becomes
zero. The values of $B_{\rm c}$ depend exclusively on the geometry
of the considered setup. The field strength is in units of $B_0$.}
\label{fig:4}
\end{figure}
\end{center}

It is known that in the normal case the linear-response
low-temperature conductance is simply proportional to the
transmission coefficient for electrons $T_{\rm e}$, according to
Landauer's formula $G_{\rm N}=(2e^2/h)\;T_{\rm e}$. Lambert et
al.~\cite{Lam93} have worked generalizations for systems including
superconducting islands or leads. For the Andreev version of the
Sinai billiard system, the conductance is given by $G_{\rm
S}=(2e^2/h)\;(T_{\rm e}+R_{\rm h})$ where $R_{h}$ is the
reflection coefficient for holes. In analogy with the quantum
mechanical case, we plot in figure~\ref{fig:4} the
magnetoconductance of a normal (open circles) and a
superconducting (filled stars) antidot for the four setups of the
Sinai billiard considered, using the above formulae.

From figure~\ref{fig:4} we see that for all the setups considered,
with increasing field strength the dependency of the classical
trajectories on the applied magnetic field drives the classical
dynamics from mixed to regular for both versions of billiards.
This is grossly reflected in the non-monotonic behavior of the
magnetoconductance, in agreement with the behavior already
observed in figure~\ref{fig:3} for the transmission and reflection
probabilities.

Some general comments are in order: At non vanishing external
field the classical dynamics of both the normal and Andreev
billiards is characterized by a mixed phase space of coexisting
regular and chaotic regions. At $B=0$ the superconducting antidot
leads to an integrable dynamics since trajectories are precisely
retraced after retroreflection while the corresponding normal
device possesses a mixed phase space. There are three families of
periodic orbits each forming a continuous set that occur in the
classical dynamics and phase space of the closed
system~\cite{Gaspard,Kovacs,Silva}, i.e. without leads, leaving
their fingerprints in the open system with the attached leads. We
will briefly discuss these periodic orbits in the following. At
zero field there are orbits bouncing between two opposite walls
with velocities parallel to the normal of the corresponding walls.
At finite but weak $B$-field strength the periodic orbits form a
rosette and incorporate collisions with the antidot and the walls.
These periodic orbits are typical, i.e. dominant up to a critical
field value $B_{\rm c}$. For magnetic fields above $B_{\rm c}$ the
cyclotron radius is so small that no collisions with the antidot
can occur and skipping orbits, describing the hopping of the
electrons along the billiard walls, become dominant. The values of
this critical field $B_{\rm c}$ are marked in the relevant plots
by the dotted lines. All periodic orbits possess an eigenvalue one
of their stability matrix~\cite{Fliesser} and all periodic orbits
possess unstable directions. We remark that the above-discussed
periodic orbits of the closed billiard are not trajectories
emerging from and ending in the leads of the open billiard.
However, trajectories of particles coupled to the leads (i.e.,
injected and transmitted/reflected) can come close to the periodic
orbits of the open billiard thereby tracing their properties. This
way the presence of the periodic orbits reflects itself in the
transport properties.

Overall, we see that that in the presence of Andreev reflection
the conductance of the system is larger than in the
normalconducting case for magnetic fields $B<B_{\rm c}$. This
holds for the setups (a), (b) and (c) while for the setup (d) we
see that $G_{\rm N}\sim G_{\rm S}$. Turning on the
superconductivity at the Sinai-billiard disc, the interplay of the
bending of the trajectories and the occurring particle-to-hole
conversion accounts for a significant increase in the reflection
coefficient of holes   and therefore for this qualitatively
different behavior between the normal and Andreev version of the
billiard. On the other hand, setup (d) is an exceptional case,
where due to the circular billiard-shape the typical trajectories
in the normalconducting case for $B < B_{\rm c}$ give strong
contribution to the process of electron transmission. The same
orbits, say in setup (a) would contribute to the process of
electron reflection, spanning the difference $G_{\rm N}-G_{\rm
S}$. For $B > B_{\rm c}$ now, we expect a similar behavior of
$G_{\rm N}$ and $G_{\rm S}$ as discussed previously. The small
deviation that appears in setup (b) is due to the fact that the
interesting feature of the even number of collisions with the
antidot which is related to the generic properties of Andreev
reflection~\cite{F05} is destroyed and a small percentage of
trajectories showing an odd number of collisions with the
circumference of the disc exist, even in the region beyond the
critical value of $B$. Setup (c), i.e. the rectangular cavity
formed by reducing the z-axis boundary length, gives, as we may
see from figure~\ref{fig:4}, the smaller $G_{\rm N(S)}$ values,
with the geometry of this setup being responsible for both the
increase in the transmission of holes (and thus the reduction of
$G_{\rm S}$) in the superconducting case and the reflection of
electrons (and thus the reduction of $G_{N}$) in the corresponding
normalconducting case.

We performed simulations of the classical dynamics of low-energy
(quasi)particles and identified the magnetoconductance spectrum of
four different geometrical setups emerging out of the general
geometry of the Sinai billiard shown in figure~\ref{fig:1}. For
each setup, we studied both the normal and Andreev version of the
Sinai billiard, i.e. we investigated the interplay between
trajectory bending and Andreev reflection and showed how such
effects influence the overall (magneto)transport properties of
Andreev billiards when compared to their normal counterparts. The
classical simulations reported here are not severely demanding in
computer time and can be easily tuned according to the parameters
defining the setup, i.e. the shape of the cavity, the
position/size of the scattering disc and the position/width of the
leads. Therefore, we envisage that our study could be further
developed and utilized both theoretically and experimentally in
future investigations.

{}


\begin{thebibliography}{}

\bibitem{QuaCh} K.-F. Berggren, S. {\AA}berg (eds), QUANTUM CHAOS Y2K: Proceedings
of Nobel Symposium 116, World Scientific, 2000.

\bibitem{Alha} Y. Alhassid, Rev. Mod. Phys. {\bf 72}, 895 (2000).

\bibitem{Ri00} K. Richter, Semiclassical Theory of Mesoscopic Quantum Systems, in
vol. 161 of Springer Tracts in Modern Physics, Springer, Berlin,
2000.

\bibitem{Jalab} R.A. Jalabert, in New Directions in Quantum Chaos, edited by G.
Casati, I. Guarneri, U. Smilansky, IOS Press, Amsterdam, 2000.

\bibitem{KosMasGol} I. Kosztin, D.L. Maslov, and P.M. Goldbart, Phys. Rev. Lett. {\bf 75}, 1735 (1995).

\bibitem{S-billiards} C.W.J. Beenakker, Lect. Notes Phys. {\bf 667}, 131 (2005).

\bibitem{EPL02ETW} J. Eroms, M. Tolkiehn, D. Weiss, U. R\"{o}ssler, J. DeBoeck, and S.
Borghs, Europhys. Lett. {\bf 58}, 569 (2002).

\bibitem{And64} A. F. Andreev, JETP {\bf 19}, 1228 (1964).

\bibitem{MBFC96} J. Melsen, P. Brouwer, K. Frahm, and C. Beenakker, Europhys. Lett. {\bf
35}, 7 (1996).

\bibitem{PRB00CBA} A.A. Clerk, P. W. Brouwer, and V. Ambegaokar, Phys. Rev. B  {\bf 62}, 10226 (2000).

\bibitem{EPJB01ILV} W. Ihra, M. Leadbeater, J. L. Vega, and K. Richter, Eur. Phys. J. B {\bf
21}, 425 (2001).

\bibitem{PRL99SB} H. Schomerus and C.W.J. Beenakker, Phys. Rev. Lett. {\bf 82}, 2951 (1999).

\bibitem{TA01} D. Taras-Semchuk and A. Altland, Phys. Rev. B {\bf 64}, 014512 (2001).

\bibitem{PRB04CPP} J. Cserti, P. Polin{\'a}k, G. Palla, U. Z\"{u}licke, and C.J. Lambert,
Phys. Rev. B {\bf 69}, 134514 (2004).

\bibitem{FTPR05} G. Fagas, G. Tkachov, A. Pfund, and K. Richter, Phys. Rev. B {\bf 71}, 224510 (2005).

\bibitem{F05} N.G. Fytas, F.K. Diakonos, P. Schmelcher, M. Scheid, A. Lassl, K.
Richter, and G. Fagas, Phys. Rev. B {\bf 72}, 085336 (2005).

\bibitem{Lam93} C.J. Lambert, V.C. Hui, and S.J. Robinson, J. Phys. C {\bf 5}, 4187 (1993).

\bibitem{Ketz} A.S. Sachrajda, R. Ketzmerick, C. Gould, Y. Feng, P.J. Kelly, A. Delage, and Z. Wasilewski, Phys. Rev. Lett. {\bf 80}, 1948 (1998).

\bibitem{Gaspard} P. Gaspard and J.R. Dorfman, Phys. Rev. E {\bf 52}, 3525 (1995).

\bibitem{Kovacs} Z. Kov\'{a}cs, Phys. Rep. {\bf 290}, 49 (1997).

\bibitem{Silva} L.G.G.V. Dias Da Silva and M. A. M. de Aguiar, Eur. Phys. J. B {\bf 16}, 719 (2000).

\bibitem{Fliesser} M. Fliesser, G. J. O. Schmidt, and H. Spohn, Phys. Rev. E {\bf 53}, 5690 (1996).

\end{thebibliography}
\end{document}